\documentclass[fleqn,oldversion]{aa}
\title{Coagulation and Fragmentation in molecular clouds}
\subtitle{II. The opacity of the dust aggregate size distribution}
\author{C.W. Ormel\inst{1} \and M. Min\inst{2} \and A.G.G.M. Tielens\inst{3} \and C. Dominik\inst{4} \and D. Paszun\inst{4}}
\institute{Max-Planck-Institut f\"ur Astronomie, K\"onigstuhl 17, 69117, Heidelberg, Germany;
          \email{ormel@mpia-hd.mpg.de}
          \and
          Astronomical Institute Utrecht, Utrecht University, P.O. Box 80000, 3508 TA Utrecht, The Netherlands
          \and
          Leiden Observatory, Leiden University, P.O. Box 9513, 2300 RA Leiden, The Netherlands\\
          \email{tielens@strw.leidenuniv.nl}
          \and
          Sterrenkundig Instituut ‘Anton Pannekoek’, Kruislaan 403, 1098 SJ Amsterdam, The Netherlands; 
          \email{C.Dominik@uva.nl}
          }
\usepackage{natbib,graphicx}
\usepackage{amsmath,amsbsy}
\usepackage[varg]{txfonts}

\newcommand{\se}[1]{\mbox{Sect.\ \ref{sec:#1}}}

\newcommand{\Se}[1]{\mbox{Section\ \ref{sec:#1}}}
\newcommand{\eq}[1]{\mbox{Eq.\ (\ref{eq:#1})}}

\newcommand{\fg}[1]{\mbox{Fig.\ \ref{fig:#1}}}
\newcommand{\fgs}[2]{Figs.\ \ref{fig:#1} and \ref{fig:#2}}
\newcommand{\Fg}[1]{\mbox{Figure\ \ref{fig:#1}}}
\newcommand{\Tb}[1]{\mbox{Table\ \ref{tab:#1}}}

\newcommand{\ie}{i.e.,}
\newcommand{\vs}{vs.}
\newcommand{\eg}{e.g.,}

\newcommand{\cf}{cf.}
\newcommand{\simu}[1]{\texttt{#1}}
\newcommand{\micr}{\ensuremath{\mu\mathrm{m}}}

\abstract{
The dust size distribution in molecular clouds can be strongly affected by ice-mantle formation and (subsequent) grain coagulation. Following previous work where the dust size distribution has been calculated from a state-of-the art collision model for dust aggregates that involves both coagulation and fragmentation (Paper I), the corresponding opacities are presented in this study.  The opacities are calculated by applying the effective medium theory assuming that the dust aggregates are a mix of 0.1\ \micr\ silicate and graphite grains and vacuum. In particular, we explore how the coagulation affects the near-IR opacities and the opacity in the 9.7 \micr\ silicate feature.  We find that as dust aggregates grow to \micr-sizes both the near-IR color excess and the opacity in the 9.7 \micr\ feature increases.   Despite their coagulation, porous aggregates help to prolong the presence of the 9.7 \micr\ feature.  We find that the ratio between the opacity in the silicate feature and the near-IR color excess becomes lower with respect to the ISM, in accordance with many observations of dark clouds.  However, this trend is primarily a result of ice mantle formation and the mixed material composition of the aggregates, rather than being driven by coagulation.  With stronger growth, when most of the dust mass resides in particles of size $\sim$10\ \micr\ or larger, both the near-IR color excess and the 9.7 \micr\ silicate feature significantly diminish.  Observations at additional wavelengths, in particular in the sub-mm range, are essential to provide quantitative constraints on the dust size distribution within dense cores. Our results indicate that the sub-mm index $\beta$ will increase appreciably, if aggregates grow to $\sim$100 \micr\ in size.
}
\authorrunning{C.W. Ormel et al.}
\newcommand{\figw}{0.45\textwidth}
\begin{document}
\maketitle
\section{Introduction}
Interstellar grains are an important component of molecular clouds. Interstellar dust is a major source of opacity in the far-ultraviolet region of the spectrum and hence the dust characteristics are an important aspect of the spectral energy distribution of molecular clouds. In addition, through their influence on photodestruction rates, dust controls the survival of molecules \citep{RobergeEtal1991}.  Grain surfaces also provide a `meeting' place for atomic and molecular species and hence enable reactions to proceed efficiently \citep{TielensHagen1982,HasegawaEtal1992}. Thus, molecular hydrogen is generally thought to be formed `exclusively' on grain surface \citep{HollenbachSalpeter1971} and the formation of other species such as water may also well be dominated by grain surface chemistry \citep{CharnleyEtal1992,HasegawaEtal1992}. Finally, grains are often used as a proxy for the gas and observations at, in particular, sub-millimeter wavelengths provide a convenient probe of the density distribution of clouds \citep{JohnstoneBally2006}. Such observations are therefore often employed in analyzing the star formation characteristics of molecular clouds in terms of, for example, the masses of cloud cores and protostellar condensations \citep{AlvesEtal2007,JorgensenEtal2008}. The physical characteristics of interstellar dust are therefore key to understanding many aspects of molecular clouds and their evolution.

The properties of interstellar dust, however, evolve inside dense clouds \citep{Dwek1998,ZhukovskaEtal2008}. Observationally, large grains are indicated inside dense clouds through an increase in the value of the total-to-selective extinction ratio, $R_V$, from about 3.1 in the diffuse interstellar medium to values as high as 5.5 for sight lines traversing dense cores \citep{WilkingEtal1980,Whittet2005,OlofssonOlofsson2010}. This is supported by studies of the wavelength dependence of polarization which also indicates grain growth \citep{CarrascoEtal1973}. The unusually low value of the visual extinction per H-atom towards the dense cloud, $\rho$-Oph implies that grain growth -- at least for this cloud -- reflects coagulation rather than growth through mantle formation (Jura 1980).  Early studies have revealed that for low column densities in the diffuse ISM, the near- and mid-IR extinction shows a `universal' behavior characterized by a power-law dependence of extinction on wavelength with an exponent of $-1.8$ \citep{RiekeLebofsky1985,MartinWhittet1990}. Recently, large scale surveys at near- and mid-IR wavelength (2MASS, UKIDSS and GLIMPSE) have provided new insights into the extinction behavior of dust deep inside dense clouds by probing the colors of background stars.  These studies have revealed that the near-IR extinction law shows a flattening in dense cores, typically at depths corresponding to an $A_V$ of $\sim$20 magnitudes \citep{IndebetouwEtal2005,ChapmanEtal2009,CambresyEtal2011} indicative of grain growth \citep{WeingartnerDraine2001}. Recently, too, near-IR scattered light has been detected from dense cloud cores and this is indicative of grain growth to micron-sizes \citep{PaganiEtal2010,SteinackerEtal2010}. Finally, grain growth in dense clouds is also supported by an analysis of the sub-millimeter emission \citep{BianchiEtal2003,KramerEtal2003,ShirleyEtal2011}.

The high sensitivity of the Spitzer/IRS instrument has allowed for the first time studies of the profile of the 10 \micr\ silicate feature for dust inside dense clouds by probing background stars. These studies have revealed two new aspects of the extinction behavior of dust inside dense clouds. First, while the ratio of the mid-IR extinction to the near-IR extinction increases with increasing depth (the flattening of the extinction curve mentioned above), the total extinction at 10 \micr\ does not increase; e.g., per unit near-IR extinction, the increase in the 10 \micr\ `continuum' extinction is accompanied by a concomitant decrease in the strength of the silicate feature \citep{ChiarEtal2007,McClure2009,BoogertEtal2011}. This decrease in the optical depth of the silicate band relative to the total near-IR extinction is accompanied by a shift to the blue in the profile compared to that of the diffuse ISM \citep{vanBreemenEtal2011,ChiarEtal2011}. In this study we will investigate the role of dust coagulation on the behavior of the 10 \micr\ silicate opacity \vs\ the near-IR color excess.

Variations in the dust properties inside dense clouds can reflect accretion of ice mantles and/or coagulation. Taking the measured strength of the absorption features due to interstellar ice mantles towards background sources behind dense clouds \citep{BoogertEtal2011} and the intrinsic strength of these bands, the increase in grain volume due to ice is $\simeq 2.5\times 10^{-27}$ cm$^3$ per H-atom \citep[\cf][]{TielensAllamandola1987}; a little less than the silicate dust volume ($\simeq$$3.6\times 10^{-27}$ cm$^3$ per H-atom). However, these ice mantles only increase the extinction within specific absorption modes and have little influence on the overall extinction \citep{OssenkopfHenning1994}. Moreover, because the surface area of grains is dominated by the small end of the size distribution, the mantle thickness is very limited ($<$$175$ \AA\ if all oxygen were to deplete as ice; \citealt{Draine1984}). In terms of grain size, coagulation is the more important one of the grain growth processes \citep{Ossenkopf1993,WeidenschillingRuzmaikina1994}. At low collision velocities, small dust grains will stick upon collision forming larger conglomerates. Collisions among these aggregates builds then larger and larger grains \citep[\cf][]{Ossenkopf1993}. Because the coagulation process itself as also the extinction properties depend strongly on the grain structure, much theoretical and experimental effort has focused on understanding the evolution of the porosity of collisional aggregates \citep{Ossenkopf1993,DominikTielens1997,PaszunDominik2009,BlumWurm2008,SuyamaEtal2008,WadaEtal2008,Okuzumi2009}. These studies show that initially coagulation may lead to the formation of open, fractal structures, but eventually, if collision velocities become high, compaction and then fragmentation will take over \citep{OrmelEtal2009}. As a result, the structure of the aggregates -- and therefore the extinction properties -- depend on the density and history of the cloud. Recently, we have calculated the collisional growth of interstellar grains in dense clouds \citep{OrmelEtal2009}. Here, we examine the implications of this growth for the extinction behavior of dust in dense clouds. 

This paper is organized as follows. \Se{overview} provides an overview of previous work and details the assumptions employed in the calculation of the opacities. In \se{opac-spheres} the extinction coefficients are calculate for solid grains, varying only their composition (silicates, graphites, ice-layers, and mixed variants). \se{opac-spheres} also discusses the sensitivity of the porosity of the dust aggregates.  In \se{opac-aggr}, we apply the opacity calculations to the aggregate dust size distribution obtained from the \citet{OrmelEtal2009} study. \Se{summ} discusses several implications and summarizes.

\section{\label{sec:overview}Overview of previous work and methodology}
The dust opacities that we present in this work are the result of a chain of modeling efforts. These include:
\begin{enumerate}
  \item a collisional dynamics code to perform a parameter study on the outcome of a collision between two aggregates of grains \citep{PaszunDominik2009};  
  \item a collisional evolution code to compute the size distribution of dust aggregates as function of time (\citealt{OrmelEtal2009}; henceforth Paper I);
  \item an effective medium approach to calculate the optical properties of porous dust aggregates \citep{MinEtal2008}; and
  \item a Mie scattering code.
\end{enumerate}
In the following, we briefly review these steps. 

\subsection{Collisions among dust aggregates}
At low velocity, collisions among dust grains result in loosely bound structures held together by surface forces that act on the contact points between the individual grains.  These structures are referred to as dust aggregates and subsequent `growth' involves collisions between two dust aggregates.  The outcome of a collision between two dust aggregates depends on the material properties, the internal structure of the aggregates, and on the collision properties (impact parameter, collision velocity, mass ratio).  In a pioneering work, \citet{DominikTielens1997} derived \textit{collision recipes} to predict the outcome of collisions between two arbitrary aggregates, using appropriate scaling behavior.  In particular, the outcome of a collision between two grains or aggregates -- sticking, compaction, or fragmentation -- depends on the collision energy in comparison to a critical energy and the number of contacts within the aggregate (see \citealt{BlumWurm2008} for a review). 


Using a 3D molecular dynamics code, in which the equations of motion for each individual grain within the aggregate are solved, \citet{PaszunDominik2009} have expanded the study of \citet{DominikTielens1997} by including fragmentation and the influence of porosity on the collision outcome.  Using look-up tables, the \citet{PaszunDominik2009} recipe provides a prescription for the outcome of the collision: the size distribution of the collisional fragments (when present) and of the porosity increase/decrease of the collision products.  The great advantage (for modelers) of using the recipe approach is that we only need to follow the average properties of the particles (the porosity and size of the aggregates, material properties of the grains) and the collision properties (\ie\ the velocity field), and can dispense of modeling the full substructure of dust aggregates.  Indeed, the latter approach is computationally too expensive since molecular dynamics code can only treat a limited number of grains.

\begin{figure*}
  \centering
  \includegraphics[width=0.8\textwidth,clip]{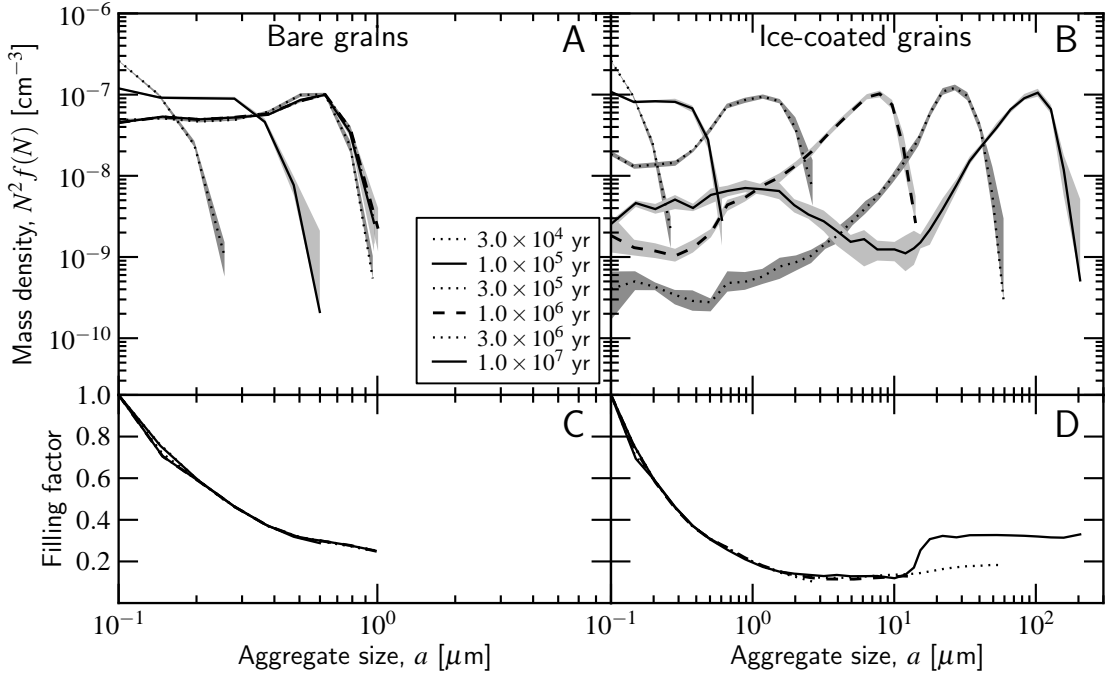}
  \caption{\label{fig:distr}The size distribution of dust aggregates as function of time as calculated by \citet{OrmelEtal2009} for bare grains (left panel) and ice-coated grains (right panel).  The gas density is $n_g=10^5\ \mathrm{cm}^{-3}$.  For bare grains the size distribution evolves towards a steady state due to the emergence of fragmentation, while aggregates with ice-coated grains keep growing. The gray shading denotes the spread obtained from independent runs of the Monte Carlo code. The lower panels show the filling factor of the dust aggregates.  Decreasing filling factors indicate a fractal structure.  But for larger aggregates compaction halts the fractal growth.  }
\end{figure*}
\subsection{Application to molecular cloud cores}
In Paper I we have included these recipes in a Monte Carlo code to compute the size distribution of dust aggregates as function of time.  In Monte Carlo codes \citep[\eg][]{OrmelEtal2007,ZsomDullemond2008} the computational particles are characterized by their properties.  In this case, the aggregates were represented by their mass $m$ and their projected surface-area over mass ratio, $\sigma/m$.   An accurate prescription for the latter property is very important, as it plays a critical role in the determination of the relative velocities among the dust aggregates.  In Paper I we explicitly follow both $\sigma$ and $m$, \ie\ $\sigma$ is not \textit{a priori} assumed to be a function of $m$ as, \eg\ for compact spheres ($\sigma \propto m^{2/3}$) or very fluffy particles $\sigma \propto m$ \citep{MinatoEtal2006}.

Equivalently, we can express $\sigma$ as a size, $a_\sigma = \sqrt{\sigma/\pi}$, and define the \textit{geometrical filling factor} as
\begin{equation}
  \phi_\sigma \equiv \frac{\textrm{Volume occupied by $N$ grains}}{\textrm{Volume equivalent to surface area}} = \frac{N\times 4\pi a_0^3/3}{4\pi a_\sigma^3/3} = \frac{Na_0^3}{a_\sigma^3},
  \label{eq:phisigma}
\end{equation}
where we have assumed that all $N$ grains are spherical and of the same size $a_0$.  This monodisperse grain approximation is used in both the parameter study of \citet{PaszunDominik2009} as well as in Paper I; in the latter study the radius of the grains $a_0$ was fixed at $0.1\ \mu$m.  


In this study we furthermore limit our results to models that start out at a gas density of $n_\mathrm{g} = 10^5\ \mathrm{cm}^{-3}$. 
\Fg{distr} shows the evolution of size distribution for this density.  Here, $f(N)$ gives the number density spectrum of aggregates by their number of grains, $N$.  Following Paper I, we multiply by $N^2$ to obtain a proper comparison for the mass density for a logarithmic scaling.  We expect that for other gas densities the coagulation will follow a similar trend (at least initially) but at different timescales (see Paper I). Paper I considered two modes of aggregation: aggregation among \textit{bare} (silicate) grains and among \textit{ice-coated} grains.  For the sticking behavior, it is the surface properties that matter and even a modest amount of freeze out on bare (silicate or carbonaceous) grains can significantly speed-up the coagulation, since the critical energies are expected to be much larger \citep{DominikTielens1997,WadaEtal2009}.  We indeed observed a rather sensitive dependence of our results on these material properties.  In the bare silicate models the aggregates were weak and were prone to fragmentation.  Consequently, a steady state between coagulation and fragmentation with little net growth was quickly established on timescales larger than $\sim$$3\times10^5$ yr (\fg{distr}a).  However, if the grains were coated by a (thin) ice layer, small grains were quickly swept into aggregates.  These aggregates continued to grow through mutual collisions.  Initially, the aggregates remained rather porous as the collisional energies were low, but with increasing size and increasing energy the aggregates started to compact.  Only on timescales $\gtrsim$$10^7$ yr were collisional energies powerful enough to provide some form of erosion and replenish the smaller grains (\fg{distr}b).

In \fg{distr}c,d the (mean) filling factor (\eq{phisigma}) of the aggregates is plotted, as function of their size.  Most of the curves fall on top of each other, indicating that $\phi_\sigma$ is a function of size only.  Initially, the filling factor decreases with size, which indicates that the growth is fractal.  Due to subsonic turbulent motions aggregates move at an appreciable relative velocity of $\approx$10 m s$^{-1}$ which increases with the growth of these particles.  As a result, energetic collisions will halt the decrease in $\phi_\sigma$ through compaction.  After $t=10^7$ yr there is a noticeable increase in $\phi_\sigma$ towards $\phi_\sigma = 33\%$, which is the maximum adopted value of $\phi_\sigma$ for evolved aggregates (\citealt{GuettlerEtal2009}, Paper I). 


\subsection{Optical properties of dust aggregates}
From the size distributions of aggregates provided by Paper I we compute opacities, following the method described in \citet{MinEtal2008}. \citet{MinEtal2008} describe an efficient method to obtain the optical properties of porous dust aggregates using an effective medium approach.  This involves applying a mixing rule that appropriately averages the optical constants of the various components out of which the aggregate is constituted.  The resulting effective refractive index ($m_\mathrm{eff}$) can then be used in Mie's solution of light scattering.  The \citet{MinEtal2008} study includes a prescription to add vacuum as a separate component to the mixing rule. Here we apply the method to spherical grains.  We consider the \citet{Bruggeman1935} mixing rule, with $N_c$ components and vacuum:
\begin{equation}
  f_\mathrm{fill} \sum_{i=1}^{N_c} f_i \alpha_c (m_i/m_\mathrm{eff})  +(1-f_\mathrm{fill}) \alpha_c (1/m_\mathrm{eff}) = 0,
  \label{eq:Brugg}
\end{equation}
where $m_\mathrm{eff}$ is the effective refractive index, $f_\mathrm{fill}$ the filling factor of the aggregates (see below), $f_i$ the volume fraction of component $i$ with $\sum_i^{N_c} f_i = 1$, $m_i$ the refractive index of component $i$, and $\alpha_c(m)$ the polarizability of the components.  For homogeneous spheres, the polarizability is  
\begin{equation}
  \alpha_\mathrm{hs}(m) = 4\pi a^3 \frac{m^2-1}{m^2+2},
  \label{eq:alpha-hs}
\end{equation}
where $a$ is the aggregate's radius. For ice-coated spheres, the polarizability becomes \citep{VanDeHulst1981}
\begin{equation}
  \alpha_\mathrm{cs}(\varepsilon) = 4\pi a^3 \frac{(\varepsilon_2-1)(\varepsilon_1+2\varepsilon_2) + Q^3(2\varepsilon_2+1)(\varepsilon_1-\varepsilon_2)}{(\varepsilon_2+2)(\varepsilon_1+2\varepsilon_2) + Q^3(2\varepsilon_2-2)(\varepsilon_1-\varepsilon_2)},
  \label{eq:vdHulst}
\end{equation}
where $\varepsilon_1 = m_1^2$ is the dielectric constant of the inner core, $\varepsilon_2$ the dielectric constant of the coated material, and $1-Q$ the fractional size of the coated material;  that is, the inner core consists of fractional volume $Q^3$.  In our case, $\varepsilon_1$ represents silicates or graphite and $\varepsilon_2$ ices, where we take $Q=0.9$ for a thin coating. The refractive indices are taking from \citet{Draine2003}.

\begin{figure}
  \includegraphics[width=\figw,clip]{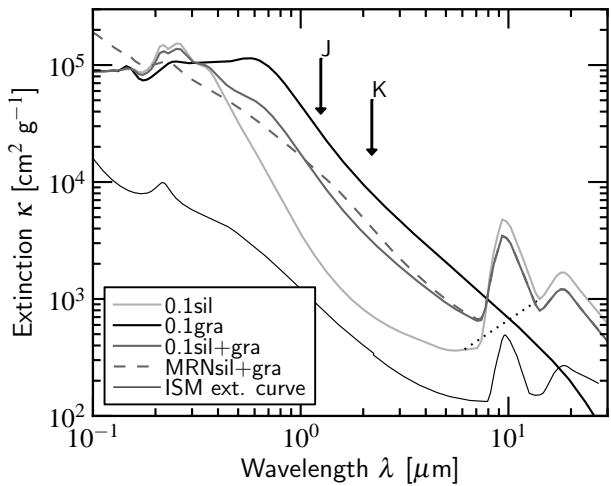}
  \caption{\label{fig:spec0}The near/mid-IR extinction for spherical grains.  The curves show the opacities corresponding to 0.1 \micr\ spheres of silicate (light gray) and graphite (black).  Dark gray curves correspond to a 1:2 mix of graphite to silicates of 0.1 \micr\ grains (solid gray curve) and for the MRN distribution (dashed curve).  The dotted curve shows the adopted continuum for calculating $\kappa_\mathrm{sil}$, the amount of extinction in the 9.7 \micr\ silicate feature. Shown for illustrative purposes is the observed ISM extinction curve (thin solid curve) with arbitrary scaling in the $y$-direction.}
\end{figure}
Thus, \eq{Brugg} is solved for $m_\mathrm{eff}$ (or $\varepsilon_\mathrm{eff}$). In general, this requires us to solve a polynomial equation, which can be done numerically or analytically if the degree is low. The only other remaining parameter is the filling factor $f_\mathrm{fill}$.  \citet{MinEtal2008} suggests to use the radius of gyration of the aggregate $a_g$, \ie\ $f_\mathrm{fill} = N a_0^3/a_g^3$, and reports an excellent match with more rigorous coupled dipole approximation calculations.  However, from Paper I only $a_\sigma$ (rather than $a_g$) is available, since we have not computed the full substructure of the aggregates.  For this study we will assume that these radii are similar, $a_\sigma \approx a_g$.  For sufficiently dense aggregates of fractal dimension $D_F>2$, one can expect that the projected surface area is indeed similar to, although always smaller than, the outer-most radius (see Paper I for a quantitative relation between these radii).  The same arguments holds for the gyration radius $a_g$.  For fluffy aggregates of $D_F<2$ these assumptions will fail, however, and our method becomes invalid. But in our case aggregates never become very fluffy ($\phi$ always exceeds 10\%, see \fg{distr}). Therefore, these radii are similar, which allows us to substitute $\phi_\sigma$ for $f_\mathrm{fill}$. 

From the effective optical constant $m_\mathrm{eff} = n_\mathrm{eff} + ik_\mathrm{eff}$ and a particle of effective size $a_\sigma$, we apply the Mie scattering solution to Maxwell's equations \citep[\eg][]{BohrenHuffman1983}.  We then obtain the scattering cross section $C_\mathrm{sca}(\lambda)$, the absorption cross section $C_\mathrm{abs}$, and the extinction cross section, $C_\mathrm{ext} = C_\mathrm{sca}+C_\mathrm{abs}$, as function of wavelength $\lambda$.  From these cross sections and the particle distribution function we finally obtain the mass-extinction coefficient, $\kappa_\mathrm{ext}(\lambda)$, which gives the cross section for extinction per unit mass dust.

\section{Opacities for static grain distributions}
\label{sec:opac-spheres}
\subsection{The near-IR extinction law}
\label{sec:nir-ext}
\begin{table*}
  \begin{center}
  \caption{\label{tab:list}Model runs.}
  \begin{tabular}{lllll}
    \hline
    \hline
    Name      & Mixing type & Size distribution   &  Components         &  Figure ref.  \\
    (1)      & (2)      & (3)                 & (4)                     &  (5) \\
    \hline
    0.1sil        &          & Spheres, 0.1 \micr  & silicate           & \fgs{spec0}{rplot0} \\
    0.1gra        &          & Spheres, 0.1 \micr  & graphite           & \fgs{spec0}{rplot0} \\
    0.1sil+gra    & I        & Spheres, 0.1 \micr  & silicate,graphite  & \fgs{spec0}{rplot0} \\
    MRNsil+gra    & I        & Spheres, MRN        & silicate,graphite  & \fgs{spec0}{rplot0}\\
    0.1ic-sil     &          & Spheres, 0.1 \micr  & ic-silicate  & \fg{rplot0}\\
    0.1ic-gra     &          & Spheres, 0.1 \micr  & ic-graphite  & \fg{rplot0}\\
    0.1ics+icg    & I        & Spheres, 0.1 \micr  & ic-silicate,ic-graphite  & \fg{rplot0} \\
    0.1(sil,gra)  & II       & Spheres, 0.1 \micr  & silicate,graphite  & \fg{rplot0}\\
    \hline
    (sil,gra)       & II     & Aggregates          & silicate, graphite   & \fg{specs}a \\
    (ic-sil,ic-gra) & II     & Aggregates, coated  & ic-silicate, ic-graphite & \fg{specs}b \\
    (ic-sil,gra)    & II     & Aggregates, coated  & ic-silicate, graphite    & \fg{specs}c \\
    ic-sil+gra      & I      & Aggregates          & graphite,ic-silicate                 & \fg{specs}d\\
    \hline
  \end{tabular}
  \end{center}
  Note.|Overview of all model runs.  (1) Model name. The top part of the table corresponds to the static grain distributions conducted in \se{opac-spheres}, whereas the bottom rows concern the opacity calculation of the (evolving) aggregate distributions from \citet{OrmelEtal2009} (\se{opac-aggr}).  (2) The considered mixing type (if applicable), see \fg{mixing}.  Type I indicates the two components (Col.\ 4) are mixed in a spatial volume; type II indicates they are mixed within aggregates or grains.  (3) Adopted size distribution.  For the static grain distributions these are either monodisperse 0.1 \micr\ grains or follow an MRN distribution.  For the aggregate models these corresponds to either \fg{distr}a (non-coated) or \fg{distr}b (ice-coated), respectively. (4) Components included in the aggregate calculations.  Here, `ic' stands for ice-coated. The vacuum component that is present in the aggregate models is not listed.  In every model where both silicates and graphites are present their mass ratio is fixed at 2:1. 
\end{table*}
\Tb{list} gives an overview of all opacity calculations.  We first consider opacities resulting from individual grains, or grain mixtures.  In \fg{spec0} the light gray curve gives the mass-weighed extinction coefficient for 0.1\ \micr\ silicate grains, the black curve correspond to graphite grains, and the dark gray curve corresponds to a silicate-graphite mix of grains in a 2:1 ratio (we motivate this choice for the ratio below in \se{qval0}). The 9.7 \micr\ feature is prominently visible in the silicate curve and in the silicate-graphite mix: the silicates are the carrier for the 9.7 \micr\ feature.  At near-IR wavelengths the carbonaceous grains dominate the opacity;  they are the carriers for the $E(J-K_s)$ extinction.  

For comparison, we also plot in \fg{spec0} the observed ISM extinction curve (thin solid line). At short wavelengths we adopt the \citet{CardelliEtal1989} fit (with coefficients from \citealt{Fitzpatrick1999} and $R_V=3.1$), whereas at wavelengths $>$$2.2$ \micr\ we adopt the profile of \citet{ChiarTielens2006}.  The fit plotted in \fg{spec0} provides $A(\lambda)$ multiplied by an arbitrary constant (only the shape is meaningful). The comparison with the observed ISM extinction curve is meant to be for illustrative purposes; it is not our intention to actually \textit{fit} this curve.

We have also calculated opacities for an MRN \citep{MathisEtal1977} distribution of grains, where the grain size $a$ is distributed according to a $-3.5$ power-law between a lower size of 50 \AA\ and an upper size of 0.25\ \micr.  
The extinction coefficients are again calculated for a 2:1 silicate:graphite mix (dashed gray curve in \fg{spec0}).  Comparing $\kappa_\mathrm{ext}$ to the 0.1 \micr\ mix (solid gray curve) one recognizes minor differences.  This is understandable since in the Rayleigh limit $\kappa_\mathrm{ext}$ depends only on the mass of the grains.  However, for $\lambda \lesssim 0.1$ \micr\ the opacities of the MRN distribution become larger than those of the 0.1 \micr\ grains due to their larger total surface area per unit mass.

\begin{figure}
  \centering
  \includegraphics[width=\figw,clip]{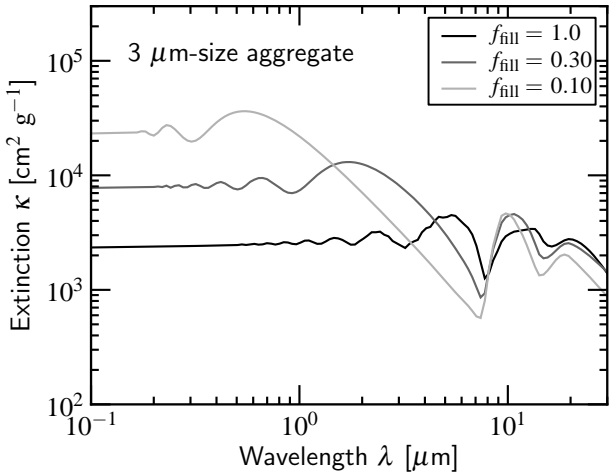}
  \caption{\label{fig:aggr}Opacity changes due to porosity variation (indicated by the filling factor $f_\mathrm{fill}$) for a silicate aggregate of size 3 \micr.}
\end{figure}
In \fg{aggr} we illustrate the opacities resulting from the Bruggeman mixing rule \eq{Brugg} for a $3$ \micr\ silicate aggregate of different porosity, indicated by the filling factor $f_\mathrm{fill}$.  The filling factor of the aggregates reaches a minimum of $\approx$10\% (see \fg{distr}).  As explained above these are porous particles, but in no way resemble very fluffy aggregates of low fractal dimension.  However, \fg{aggr} shows that the effect on the opacities is not negligible, especially at small wavelengths ($\lambda \ll 1$ \micr).  For a filling factor of 10\% the extinction is increased by a factor of 10.  This effect can be understood from geometric considerations: the surface area per unit mass is increased by a factor 10.  Moreover, the $\sim$10\ \micr\ spectral feature is more pronounced than in the low porosity case.  Porous aggregates preserve to a large extent the spectral signature of their constituent grains \citep{MinEtal2008}.  This also holds for the near-IR wavelengths; the $f_\mathrm{fill}=0.1$ aggregates are optically more pristine.

\subsection{The silicate 9.7\ \micr\ and $E(J-K)$ indicators}
\label{sec:qval0}
\begin{figure}
  \includegraphics[width=\figw,clip]{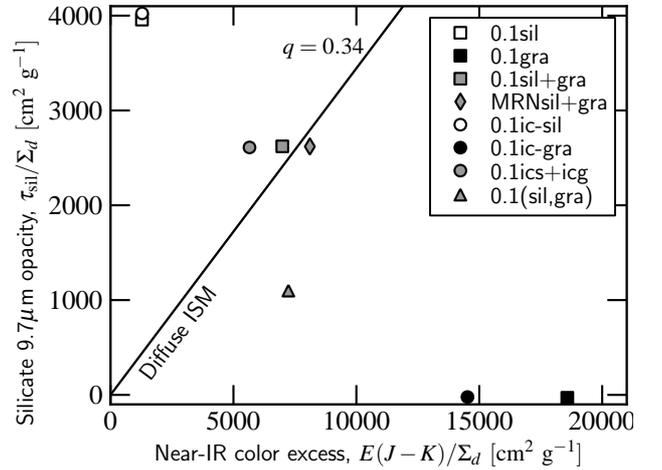}
  \caption{\label{fig:rplot0}The opacity in the 9.7\ \micr\ silicate absorption feature \vs\ near-IR color excess for spherical grains.  The colors represent the material: open symbols represent silicates, black symbols represent graphite, and gray models represent a mix of 2/3 silicate and 1/3 graphite grains.  Models involving ice-coated grains are indicated by circles and the diamond indicate an MRN distribution of spherical grains. The \texttt{0.1(sil,gra)} model denotes the opacities of a 0.1\ \micr\ sphere that consist for 2/3 out of silicates and 1/3 of graphite. The solid line of slope $q=0.34$ corresponds to the diffuse ISM. $\Sigma_d$ is the total dust column in units of \mbox{g\ cm$^{-2}$}.}
\end{figure}
We will now introduce two indicators that are used to assess the evolutionary state of a core.  These are the strength in the 9.7\ \micr\ silicate feature, obtained after the continuum subtraction (see \fg{spec0}) and the near-IR color excess, $E(J-K) = A_J - A_{Ks}$, with $A_J$ and $A_{Ks}$ the extinctions in terms of magnitudes in the $J$ (1.25 \micr) and the $K_s$ (2.2 \micr) bands, respectively.  These quantities are plotted against each other in \fg{rplot0}, where we applied the conversion between extinction in terms of optical depth or magnitudes, $A_i=1.078\tau_i$.  In \fg{rplot0} the extinction is normalized by the total dust column, $\Sigma_d$ (units: g\ cm$^{-2}$); to obtain the total extinction ($\tau_\mathrm{sil}$ or $E(J-K)$) for a particular core one must multiply by its $\Sigma_d$ value.  Clearly, for the pure silicate model (open square) the near-IR color excess is negligible, whereas for the pure graphite model (black square) the optical depth in the 9.7 feature is unimportant.  For diffuse ISM clouds, observations of the silicate optical depth and the near-IR color excess all collapse on a straight line with constant slope, $q\equiv \tau_\mathrm{sil}/E(J-K_S) \approx 0.34$ \citep{RocheAitken1984,Whittet2003,ChiarEtal2007}: the diffuse ISM curve.  The slope therefore measures the underlying opacity ratio between the indicators -- a function of the grain size distribution and composition -- which may be assumed constant in case of the diffuse ISM.

To reproduce the diffuse ISM curve the silicate and carbonaceous materials have to be mixed.  We find that with a 2:1 mix of silicates to graphite the opacity ratio falls approximately on the diffuse ISM curve (gray symbols).   Therefore, we have adopted this ratio in all our calculations. In \fg{rplot0} we also plot the indicator-values for ice-coated grains (circles) for a coating of 10\% in radius.  This has the effect of reducing the strength of the $E(J-K_s)$ colors by $\approx$25\%.  Therefore, an ice-coating can causes the indicators to deviate from the ISM value.

\begin{figure}
  \centering
  \includegraphics[width=0.45\textwidth,clip]{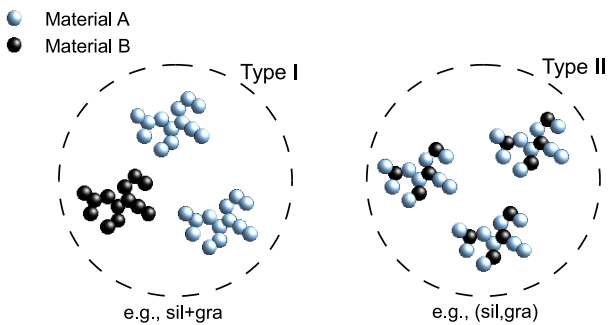}
  \caption{\label{fig:mixing}The two mixing types considered in this study. In type I mixing aggregates consisting out of grains of the same material are mixed; the opacity is calculated for each of the components individually and averaged.  In type II mixing the materials are mixed on the level of the individual aggregates, which adds another component to the Bruggeman mixing rule of \eq{Brugg}. }
\end{figure}
Finally, we have calculated the opacities for a spherical grain that consist for 2/3 of silicate and 1/3 of graphite; \ie\ we calculate the emergent opacity by applying \eq{Brugg} with $N_c = 2$ and $f_\mathrm{fill}=1$.  This mixing of the material components \textit{on the level of the individual aggregate or grain} is referred to as `type II mixing' (see \fg{mixing}) and is indicated by the brace notation in \Tb{list}, \eg\ \texttt{0.1(sil,gra)}.  In contrast, the mixing of the individual components in space that we previously encountered is referred to as type I mixing and is denoted with a `+' sign, \eg\ \texttt{0.1sil+gra}.  \Fg{mixing} provides a graphical illustration of the two mixing types, in case of aggregates.  From \fg{rplot0} it is seen that for type II mixing the emergent opacities are lower than for type I, mixing.  In type I mixing the optical properties of the participants are conserved individually and the mixed optical properties are a linear average of the individual opacities.  However, in the effective medium approach used with type II mixing (\eq{Brugg}) no such linearity is present!  In addition, the continuum subtraction applied to determine $\tau_\mathrm{sil}$ is also different between both cases.  From these results, therefore, we anticipate that the formation of aggregates composed of different materials and (to a minor extent) the formation of ice-mantles causes the indicators to deviate from the diffuse ISM curve. 

\begin{figure*}
  \centering
  \includegraphics[width=16cm,clip]{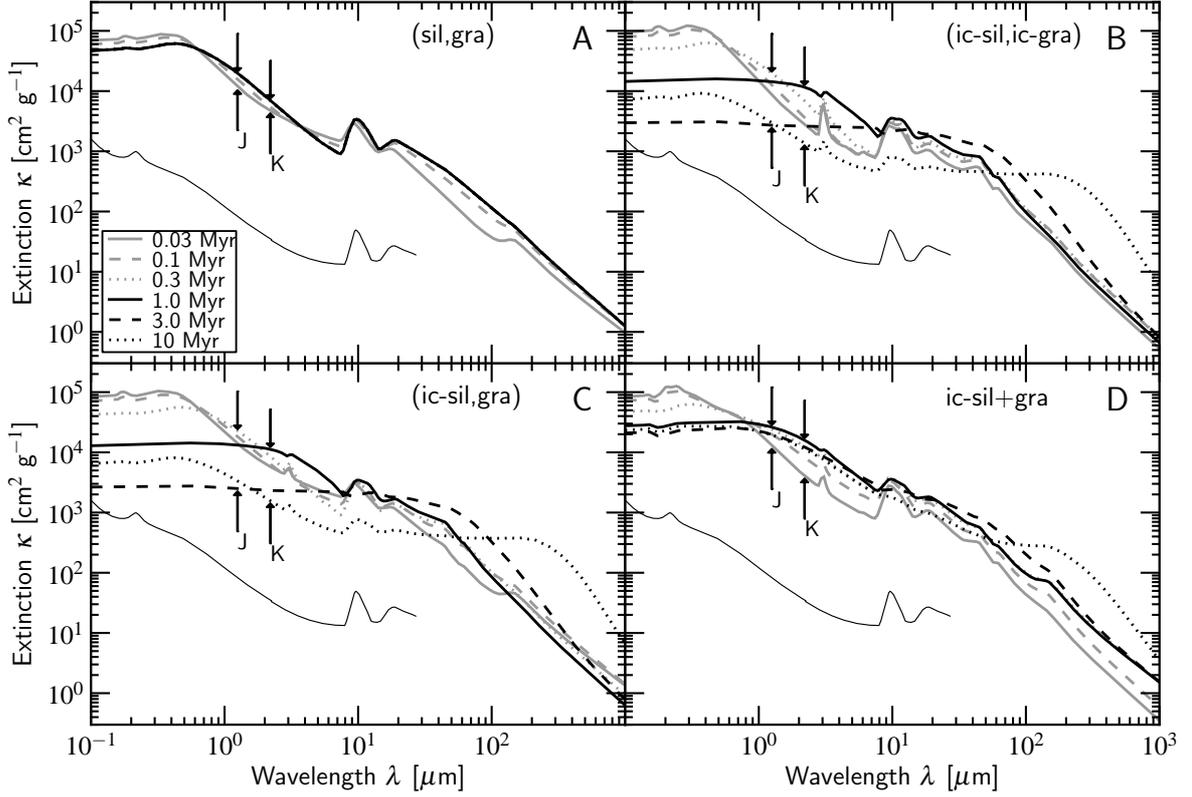}
  \caption{\label{fig:specs}The extinction for the time-dependent porous aggregates distributions at several times.  In each model silicates and graphite grains are mixed at a 2:1 ratio.  See \Tb{list} and the text for a description.  The observed ISM extinction curve (thin solid line) is shown for reference.}
\end{figure*}
\section{Opacities for evolving dust aggregates}
\label{sec:opac-aggr}
\subsection{Time-dependent extinction law} 
The bottom rows of \Tb{list} list the four aggregate models for which we have performed opacity calculations. The corresponding IR-spectra are given in \fg{specs}.  For all these models, silicates and graphites are assumed to be present at a 2:1 ratio.  In \fg{specs} the opacity distributions are shown for the same times as the size distribution in \fg{distr}. We consider four aggregate models, where we vary the size distribution (\fg{distr}a or b), and the composition of the aggregates (material components, coating).  In the first three models we assume that the aggregates consist of a mix of three materials: silicate, graphite and vacuum; \ie\ $N_c$ in \eq{Brugg} equals 2 (type II mixing).  However, in the last model we assume that the graphite and silicates are (spatially) separated and constitute different distributions.  Then, we solve for each of the $\kappa_\nu$ individually and average to obtain the mixed value (type I mixing). 

In \fg{specs}a the opacities corresponding to models without ice-coating, \ie\ as shown in \fg{distr}a, are presented (\simu{sil,gra}).  It is clear that the opacities evolve only marginally since the aggregates do not grow so much.  For $t>0.3$ Myr the curves overlap as the steady-state size distribution has been reached.  Despite the modest amount of growth, there is still a noticeable increase in $\kappa$ of a factor 2-3 for the near- and far-IR opacities.  Porous aggregates enhance the opacity at these wavelengths.

In \fg{specs}b the extinction $\kappa$ is plotted for ice-coated distributions, corresponding to \fg{distr}b.  For ice-coated models the size distribution evolves significantly (\fg{distr}b), and this is reflected in the spectra of \fg{specs}b.  Initially, for $t<1$ Myr, the trend reflects the models without coating, shown in \fg{specs}a.  The near-IR opacity significantly increases since aggregates grow to sizes comparable to these wavelengths.  Meanwhile, the opacity at wavelengths approaching $0.1$ \micr\ strongly decreases.  The opacity is especially boosted at the wavelengths that correspond to the particle size that contains most of the mass, \ie\ that peaks in \fg{distr}b.

At late times ($t>1$ Myr) this trend continues and any spectral feature is significantly reduced in strength or disappears altogether.  For the 1 and 3 Myr runs the opacity becomes gray at short wavelengths, reflecting the predominance of big, $a \gtrsim 10$ \micr, aggregates.  However, at $t=10$ Myr, there has been some replenishment of small grains due to fragmentation among aggregates (see \fg{distr}b), and this is reflected in an increase in the opacities at short wavelengths.  Also, it is clear that the sub-mm opacity has increased significantly due to the existence of $\sim$100 \micr\ aggregates (\fg{distr}b).

\subsection{Implications for $\tau_\mathrm{sil}$ \vs\ $E(J-K_s)$} 
\label{sec:implic}
In \se{qval0} we already discussed the ratio
\begin{equation}
  q \equiv \frac{\tau_\mathrm{sil}}{E(J-K_S)} = 0.93\frac{\kappa_\mathrm{sil}}{\kappa_J - \kappa_{Ks}},
  \label{eq:tau-sil:Ejk}
\end{equation}
where $\tau_\mathrm{sil}$ is the optical depth in the 9.7 silicate absorption feature (\ie\ with the continuum subtracted), $E(J-K_s)$ the near-IR color excess and $\kappa_i = \tau_i/\Sigma$ with $\Sigma$ the dust column density.  This quantity has been used as an indicator of grain growth.  Observations show that in the outer parts of dense cloud this ratio is equal to the value observed for the diffuse ISM ($q\approx 0.34$; \citealt{RocheAitken1984,Whittet2003,ChiarEtal2007}); but we can expect that grain processing (ice mantle formation and aggregation) will alter the $q$ value.

\begin{figure}
  \includegraphics[width=\figw,clip]{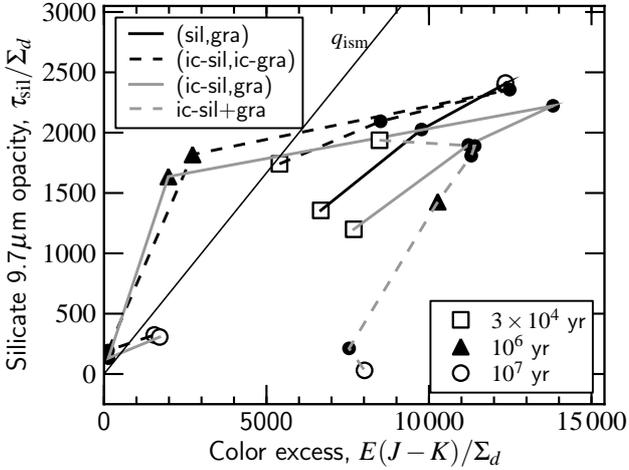}
  \caption{\label{fig:rplot1}The opacity in the 9.7 \micr\ silicate feature \vs\ near-IR color excess for the four aggregate models presented in \fg{specs}.  Each model is identified by a color/line style as indicated in the legend.  The initial ($t=0.03$ Myr) value of $\tau_\mathrm{sil}$ and $E(J-K)$ per unit column density is indicated by an empty square, the final value ($t=10$ Myr) by an empty circle, and intermediate times are indicated by black dots.}
\end{figure}
\Fg{rplot1} shows the temporal evolution of $q$ for the four aggregate models under consideration.  Per model we plot $\tau_\mathrm{sil}$ and $E(J-K_s)$ for each of the curves shown in \fg{specs} and for each of the six times, and connect these points by a line.  For the \simu{(sil,gra)} model of \fg{specs}a the points (connected by the black line) lie to the right of the ISM-curve. The $q$-value is lower than the ISM, $q\approx0.20$ -- a reduction that can be explained as the result of type II mixing (see \se{qval0}).  In the ice-coated model (\simu{(ic-sil,ic-gra)}; dashed black line in \fg{rplot1}) the $q$-value start off from the diffuse ISM line.  Initially, the behavior is very similar to the \simu{sil+gra} model: both the $E(J-K_s)$ color excess and $\tau_\mathrm{sil}$ increase and $q$ reaches a minimum value of 0.19 after $t=3\times10^5$ yr.  However, after $t>3\times10^5$ yr the effect of the ice-driven coagulation becomes prominent. At late times ($t>1$ Myr) the line approaches the origin as both indicators vanish.

Thus, the bare grain \texttt{(sil,gra)} and the ice-coated \texttt{(ic-sil,ic-gra)} models succeed in reducing $q$ below the ISM-value, which agrees with the trend of the \citep{ChiarEtal2007,ChiarEtal2011} studies.  However, it must be recognized that $q$ in these data is the net value consisting of two contributions: diffuse material from the cloud at an ISM $q$-value; and `processed' material from the core.  The observed $q$-value, then, is an upper limit to the $q$-value of the core material.  Indeed, \citet{ChiarEtal2007} suggest that the $q$ value for the core may be much less than $q_\mathrm{ISM}$.   They speculate that in the core the silicates grains (the carriers of the 9.7\ \micr\ feature) coagulate very rapidly -- essential eliminating the 10\ \micr\ feature -- whereas the carbonaceous grains (the carriers of the near-IR extinction) are not involved in this process.  Since strong growth is, within the context of our simulations, associated with ice-coated grains, this implies that the silicates succeed in acquiring ice mantles, whereas the graphite grains fail to do so.  Although the physical reason for this dichotomy is unclear, we next investigate two additional models in which the evolution of the silicates and graphites are decoupled.

First, we consider a scenario where the graphite grains are non-coated, but where the sticking behavior is still provided by the ice-coated distributions of \fg{distr}b.  In \fg{specs}c the only difference from \fg{specs}b then is that the graphite contribution present in the aggregates is not coated.  For this reason the trend in $q$ is similar.  Note that during the initial phase ($t\lesssim3\times10^5$ yr, where $q\approx0.17$) the curve very well follows the trend of \citet{ChiarEtal2011}.  Like before, this reduction is primarily the result of ice-mantle formation and aggregation (\eg\ type II mixing), rather than the very growth of these aggregates; there is no strong evolutionary trend.

Alternatively, we presume that -- again, for unexplained reasons -- silicate and graphite grains fail to interact at all.  Then, the graphite grains follow the size distribution as given in \fg{distr}a, whereas the ice-coated silicates follow \fg{distr}b.  Equivalently, we can say that the materials are spatially separated (type I mixing).  As one can see from \fg{rplot1}, this seems to be the only way to obtain low $q$ values -- provided the timescales are long enough ($\sim$Myr) to coagulate away the silicate feature.  The optical depth in the silicate feature disappears as the silicates coagulate, whereas the near-IR excess remains large, due to the presence of small graphite aggregates.
\begin{table*}
  \centering
  \caption{\label{tab:opac}Opacities at sub-mm wavelengths and indicators.}
  \begin{tabular}{lrrrrrrr}
\hline
\hline
Name  & Time &\multicolumn{4}{c}{Opacity [cm$^2$ g$^{-1}$]} & $\kappa_{850}:\kappa_{2.2}$ &  $\beta$ \\
\cline{3-6}
      &      & 150 $\mu$m& 350 $\mu$m& 500 $\mu$m& 850 $\mu$m\\
  (1) & (2) & (3) & (4) & (5) & (6) & (7) & (8) \\
\hline
(sil,gra) & $10^5$ & {$49$} & {$8.6$} & {$4.4$} & {$1.7$} & {$3.1\times10^{-4}$} & {$1.8$} \\
 & $10^6$ & {$56$} & {$9.6$} & {$4.8$} & {$1.7$} & {$2.5\times10^{-4}$} & {$1.9$} \\
 & $3\times10^6$ & {$56$} & {$9.5$} & {$4.8$} & {$1.7$} & {$2.5\times10^{-4}$} & {$1.9$} \\
 & $10^7$ & {$56$} & {$9.5$} & {$4.8$} & {$1.7$} & {$2.5\times10^{-4}$} & {$1.9$} \\
(ic-sil,ic-gra) & $10^5$ & {$50$} & {$6.8$} & {$3.2$} & {$1.1$} & {$3.0\times10^{-4}$} & {$2.0$} \\
 & $10^6$ & {$41$} & {$5.5$} & {$2.7$} & {$0.92$} & {$7.9\times10^{-5}$} & {$2.0$} \\
 & $3\times10^6$ & {$120$} & {$10$} & {$4.0$} & {$1.2$} & {$4.6\times10^{-4}$} & {$2.3$} \\
 & $10^7$ & {$410$} & {$150$} & {$58$} & {$10$} & {$7.2\times10^{-3}$} & {$3.3$} \\
(ic-sil,gra) & $10^5$ & {$60$} & {$10$} & {$5.1$} & {$1.9$} & {$3.0\times10^{-4}$} & {$1.8$} \\
 & $10^6$ & {$36$} & {$5.2$} & {$2.5$} & {$0.87$} & {$7.7\times10^{-5}$} & {$2.0$} \\
 & $3\times10^6$ & {$120$} & {$10$} & {$3.9$} & {$1.1$} & {$4.8\times10^{-4}$} & {$2.4$} \\
 & $10^7$ & {$370$} & {$170$} & {$69$} & {$12$} & {$7.3\times10^{-3}$} & {$3.4$} \\
ic-sil+gra & $10^5$ & {$42$} & {$5.3$} & {$2.5$} & {$0.86$} & {$1.2\times10^{-4}$} & {$2.0$} \\
 & $10^6$ & {$64$} & {$9.1$} & {$4.6$} & {$1.8$} & {$1.3\times10^{-4}$} & {$1.8$} \\
 & $3\times10^6$ & {$100$} & {$11$} & {$5.3$} & {$1.9$} & {$1.8\times10^{-4}$} & {$1.9$} \\
 & $10^7$ & {$290$} & {$75$} & {$29$} & {$5.9$} & {$6.3\times10^{-4}$} & {$3.0$} \\
\hline
OH94, no ice & $10^5$ & {$24$} & {$6.5$} & {$3.9$} & {$1.6$} & {$2.5\times10^{-4}$} & {$1.6$} \\
OH94, thin ice & $10^5$ & {$35$} & {$7.8$} & {$3.9$} & {$1.4$} & {$7.2\times10^{-3}$} & {$1.9$} \\
OH94 thick ice & $10^5$ & {$57$} & {$10$} & {$5.0$} & {$1.8$} & {$7.3\times10^{-3}$} & {$1.9$} \\
\hline
  \end{tabular}
  \\ Note.|Opacities at sub-mm wavelengths for the four aggregate model at several times (Cols.\ 1-6).  The bottom rows give the corresponding opacities from the \citet{OssenkopfHenning1994} models for a gas density of $n_\mathrm{gas} = 10^5\ \mathrm{cm}^{-3}$. (Col.\ 7) Ratio between the 850 \micr\ and the 2.2 \micr\ opacity. (Col.\ 8) Index of the sub-mm power law slope, $\kappa_\lambda \propto \lambda^\beta$ with $\beta$ derived from the opacity values at 850 and 500 \micr.
\end{table*}
\subsection{Opacities at sub-mm wavelengths}
Another potential powerful indicator for grain growth is the behavior at far-IR wavelengths.  However, in the Rayleigh regime ($\lambda \gg$ particle size) the opacities are not expected to be a strong function of the size of the particles.  Therefore, we see initially ($t \le 1$ Myr) only little variation among the models; and this variation can best be attributed to the way the components are mixed and whether ice-coating is involved.  \Tb{opac} provides the opacities at 150, 350, and 500 \micr\ -- the wavelengths corresponding to the SPIRE bands -- and at 850 \micr\ for the four aggregate models we considered here.  These opacities are given for several coagulation timescales.

The differences in opacity between the four aggregate models after $10^5$ yr reflect the uncertainty in the initial conditions of the dust distribution (\ie\ ice-coated or not) rather than indicating grain growth.  After $t=10^6$ yr very little has changed.  One interesting point is the different behavior seen in \fg{specs}c and d, \ie\ between the \texttt{(ic-sil,gra)} and \texttt{ic-sil+gra} models.  In the former the sub-mm opacities decrease by about a factor two, whereas in the latter there is an increase by a factor two (see also \fg{specs}).  This difference can be attributed to the applied mixing rule: the IR-opacity of free-floating graphite aggregates strongly increases with (modest) aggregation; however, if graphite is embedded in aggregates that are dominated by ice-coated silicates, its signature is suppressed. 

However, after $10^7$ yr of coagulation there is a strong increase, up to a factor 20, in the opacity of the ice-coated silicate models, which is the result of grain growth.  Most of the mass is now in 100 \micr\ size particles (see \fg{distr}b) and this boosts the emissivity at sub-mm wavelengths.  Although a factor of 5 difference in $\kappa$ may be explained by the model setup or material properties, only strong growth up to 100 \micr\ sizes can explain opacity enhancements by more than a factor of 10.

In \Tb{opac} we have compared our opacity calculation with those of \citet{OssenkopfHenning1994}.  The \citet{OssenkopfHenning1994} are based on an aggregation scenario that features only `hit-and-stick' coagulation, meaning that large fluffy (fractal) structures will form \citep{Ossenkopf1993}. This, as well as many other factors, are different from our setup. Nonetheless, after $10^5$ yr of coagulation, the sub-mm opacities are comparable, which gives support to both approaches.  However, the high opacities after $t=10^7$ yr are not present in any of the \citet{OssenkopfHenning1994} tables, also not at the higher gas densities $n_\mathrm{g}$ they consider.\footnote{Note that the most frequently quoted opacities from the \citet{OssenkopfHenning1994} study correspond to the fifth column of the opacity table in their paper (OH5), as calculated from a gas density of $10^6\ \mathrm{g\ cm}^{-3}$.  The $n_\mathrm{g}=10^5$ data in \Tb{opac} has been retrieved from the online data.}  Generally, one can expect that a higher gas density (which means a shorter collision timescale) should result in faster growth, but in the \citet{OssenkopfHenning1994} calculations, this trend is not so obvious (see also the discussion in Paper I).  

An increase in the sub-mm slope has seen recent interest with Planck observations of cold galactic clouds indicative of $\beta>2$ \citep{Planck2011,Planck2011i}. Likewise, detailed studies of individual cores \citep[\eg][]{SchneeEtal2010,ShirleyEtal2005,ShirleyEtal2011} indicate that $\beta$ is likely to be larger than 2. A favored explanation for the high $\beta$-values in these cold clumps is that the opacity becomes temperature dependent.  Several laboratory studies show that low temperatures suppress the emissivity at longer wavelengths (\citealt{MennellaEtal1998,BoudetEtal2005}; Demyk et al., in prep).  The observed correlation between $\beta$ and decreasing $T$ \citep[\eg][]{DesertEtal2008} gives credibility to this scenario. Dust aggregation is not a requirement.

On the other hand, aggregation can also change the $\beta$ value \citep{Wright1987}.  In our study, the formation of $\sim$100\ \micr\ aggregates boosts the emissivity at the short end of the sub-mm range ($\sim$100 \micr; see \fg{distr}), thereby increasing $\beta$.  Thus, the high $\beta$ results from an \textit{increase} in the sub-mm emissivity -- not a decrease, as in the low $T$ scenario. We may therefore distinguish the two scenarios by comparing the derived sub-mm opacity to the opacities at other wavelengths.  In the aggregation scenario a larger $\beta$ also results in a larger $\kappa_\mathrm{sub-mm}$:$\kappa_\mathrm{near-IR}$ ratio (see \Tb{opac}).  We do not expect this trend to emerge, however, when the increase in $\beta$ is driven by a low temperature.  In this light, it is interesting that our findings regarding the opacity ratio between 850 and 2.2 \micr\ for the (ic-sil,ic-gra) model after $3\times10^6$ yr are consistent with \citet{ShirleyEtal2011} for the B335 class 0 source. 

Polarization studies, finally, can also provide valuable insights about the typical size of aggregates.  Small aggregates can be highly anisotropic, even while the aggregation process is isotropic, which results in their alignment in a field \citep{BotetRannou2003}.  However, with increasing size the optical anisotropy is expected to diminish.

\section{\label{sec:summ}Summary and Discussion}
In this study we have investigated the effects of grain growth on the mass-weighed opacities for dense molecular clouds, focusing on IR wavelengths.  From previous works (Paper I and \citealt{MinEtal2008}) we have adopted a time-dependent size distribution and a method to calculate the opacities using an effective medium approach.  In our analysis we have mainly focused on indicators for grain growth: the ratio between the IR-colors to the strength of the 9.7 \micr\ silicate feature and the behavior at sub-mm wavelengths.  These we have attempted to link to observational studies.

In all of this, it should be emphasized that the opacity values present in this work flow from a long chain of modeling, during which several assumptions have been applied.  For example: the collisional model and the corresponding aggregate size distributions of Paper I are based on the approximation of similar-size grains (0.1\ \micr\ in this case); the gas density has been fixed at $n_\mathrm{gas}=10^5$ cm$^{-3}$; and we only considered silicate, graphite and ices as material properties.  Varying these parameters will give a different quantitative picture.  The merit of this work, then, primarily lies in the trends that are exposed when ice-mantle and aggregate formation/growth are taken into account.

Still, the current setup already provides a considerable amount of freedom to tune the opacity values.  \Fg{rplot0} shows that a considerably variation in opacity can already be achieved by merely varying the way to mix silicate and graphite grains (or their ice-coated variants), without taking into account the effects of (strong) aggregation.  These variations affect the ratio between the $9.7$ \micr\ silicate feature and the near-IR color excess, $q=\tau_\mathrm{sil}/E(J-K)$, which has been found to be constant in diffuse clouds, $q\approx0.34$, but deviating to lower values for dense cores.  Then, if we fix the ratio silicates:graphite such that $q\approx0.34$ for the diffuse ISM, we have found that an ice coating \textit{increases} $q$ if the grains are spatially mixed.  However, if the materials involved are mixed on the level of the aggregate the $q$-value substantially decreases below the diffuse ISM line.

The effects of coagulation add to these findings.  Its effect can best be understood in terms of the size of the most dominant dust aggregates in the distribution that evolve as function of time.  First, the near-IR color excess increases as aggregates grow to $\sim$\micr\ sizes.  Further growth requires grains to be ice-coated.  The near-IR color excess then first decreases as the dominant aggregate size becomes larger than the size corresponding to the J and K wavelength bands.  The strength of the 9.7 \micr\ feature decreases only thereafter when most of the dust mass becomes locked up into larger and more compact aggregates.  Thus, the finding of \citet{vanBreemenEtal2011} that for their sources it is the $J-K$ excess that affects $q$, implies, within the context of this work, that their objects are only moderately affected by coagulation.

The same conclusion may be drawn from the observed ratio of the strength in the 9.7 \micr\ feature \vs\ near-IR color excess, as presented by \citet{ChiarEtal2007,ChiarEtal2011}.  This ratio ($q$) is lower than the ISM value by roughly a factor of two, and this is in good agreement with our ice-coated models for a timescale $t<3\times10^5$ yr. We find that this decrease is, however, primarily a result of the formation of aggregates that consist of different materials, rather than their growth.  Indeed, for stronger growth, we would expect larger $q$ values to reemerge.  To obtain even lower $q$-values -- applicable to, possibly, the dense cores -- requires a decoupling of the carbonaceous and silicates, in the sense that the former species does not accrete ice mantles, whereas the latter does, and where interaction between the carbonaceous and silicate species is inhibited.  Under these conditions, our simulations show that $q\ll1$ is possible, since the carbonaceous grains only mildly coagulate, whereas the silicates aggregates become large.  Although such a scenario may seem a bit far-fetched, it is the only way to produce $q\ll1$.  It is therefore very helpful to further ascertain observationally the state of the dust in these high density cores.

In order to better constrain the evolutionary state of a cloud, multi-wavelengths observations are essential, \eg\ from the shape of the silicate feature, the scattering behavior at mid-IR wavelengths, or from sub-mm data.  As a general rule of thumb we have found that the opacity peaks for the wavelengths corresponding to the size of the dominant aggregates in the distribution.  This behavior is perhaps most obvious at sub-mm wavelengths.  Initially, little variation is expected (whereas the near-IR color excess, for example, are already strongly affected).  However, if grain growth yields aggregates of size $\sim$100 \micr, the sub-mm opacities are strongly boosted and the sub-mm power-law index $\beta$ will increase (\ie\ the opacity curve steepens).

\acknowledgements
The authors appreciate the helpful comments of the referee and editor.

\bibliographystyle{aa}
\bibliography{ads,arXiv}
\end{document}